\documentclass[aps,prd,superscriptaddress,showpacs, preprint]{revtex4-1}
\usepackage{graphicx}
\usepackage{epstopdf}
\usepackage{amsmath}
\usepackage{amsfonts}
\usepackage{amssymb}
\usepackage{latexsym}
\usepackage{hyperref}

\begin{document}
\title{Topologically Massive Spin-1 Particles and Spin-Dependent Potentials}

\author{F.A. Gomes Ferreira}\email{felipeagf@cbpf.br}

\author{P.C. Malta}\email{pcmalta@cbpf.br}

\author{L.P.R. Ospedal}\email{leoopr@cbpf.br}

\author{J.A. Helay\"{e}l-Neto}\email{helayel@cbpf.br}
\affiliation{Centro Brasileiro de Pesquisas F\'{i}sicas (CBPF), Rua Dr Xavier Sigaud 150, Urca, Rio de Janeiro, Brazil, CEP 22290-180}


\begin{abstract}
We investigate the role played by particular field representations of an intermediate massive spin-1 boson in the context of spin-dependent interparticle potentials between fermionic sources in the limit of low momentum transfer. The comparison between the well-known case of the Proca field and that of an exchanged spin-1 boson (with gauge-invariant mass) described by a 2-form potential mixed with a 4-vector gauge field is established in order to pursue an analysis of spin- as well as velocity-dependent profiles of the interparticle potentials. We discuss possible applications and derive an upper bound on the product of vector and pseudo-tensor coupling constants.
\end{abstract}


\maketitle


\section{Introduction}

Most macroscopic phenomena originate either from gravitational or electromagnetic interactions. There has been some experimental effort over the past decades towards the improvement of low-energy measurements of the inverse-square law, with fairly good agreement between theory and experiment \cite{Adelberger1}\cite{Adelberger2}. The equivalence principle has also been recently tested to search for a possible spin-gravity coupling \cite{Tarallo}. On the other hand, a number of scenarios beyond the Standard Model (BSM) motivated by high-energy phenomena predict very light, weakly interacting sub-eV particles (WISPs) that could generate new long-range forces, such as axions \cite{Wilczek}, SUSY-motivated particles \cite{Fayet} or paraphotons \cite{Okun}\cite{Holdom}\cite{Dobrescu}\cite{Accioly}. 

The discovery of a new, though feeble, fundamental force would represent a remarkable advance. Besides the Coulomb-like ``monopole-monopole" force, it is also possible that spin- and velocity-dependent forces arise from monopole-dipole and dipole-dipole (spin-spin) interactions. Those types of behavior are closely related to two important aspects of any interacting field theory: matter-mediator interaction vertices and the propagator of intermediate particles. The present paper is  mainly concerned with this issue and its consequences on the shape of the potential between two fermionic sources. This discussion is also of relevance in connection with the study, for example, of the quarkonium spectrum, for which spin-dependent terms in the interaction potential may contribute considerable corrections \cite{Bad}. Other sources (systems) involving neutral and charged particles, with or without spin, have been considered by Holstein \cite{Holstein1}.

Propagators are read off from the quadratic part of a given Lagrangean density and depend on intrinsic attributes of the fields, such as their spin. Most of the literature is concerned with spin-1 bosons in the $\left(\frac{1}{2}, \frac{1}{2}\right)$-representation of the Lorentz group (e.g., photon). Here, we would like to address the following questions: for two different fields representing the same sort of (on-shell) spin-1 particle, which role does a particular representation play in the final form of the interaction? Is the form of the mass term (corresponding to some specific mass-generation mechanism) determinant for the macroscopic characterization of the interparticle potential?

The amplitude for the elastic scattering of two fermions is sensitive to the fundamental, microscopic, properties of the intermediate boson. Our work  sets out to study the potential generated by the exchange of two different classes of neutral particles: a Proca (vector) boson and a rank-2 anti-symmetric tensor, the Cremer-Scherk-Kalb-Ramond (CSKR) field \cite{KR}\cite{CS}, mixed to another vector boson, i.e., the $\left\{ A_{\mu}, B_{\nu \kappa}\right\}-$ system with a topological mixing term. Two-form gauge fields are typical of off-shell SUGRA multiplets in four and higher dimensions \cite{Teitel1}-\cite{deWit} and the motivation to take them into consideration is two-fold:

i)	They may be the messanger, or the remnant, of some Physics beyond the Standard Model. This is why we are interested in understanding whether we may find out the track of a 2-form gauge sector in the profile of spin-dependent potentials.

ii)	In four space-time dimensions, a pure on-shell rank-2 gauge potential actually describes
 a scalar particle. However, off-shell it is not so. This means that the quantum 
fluctuations of a rank-2 gauge field may induce a new pattern of spin-dependence. Moreover, its mixing 
with an Abelian gauge potential sets up a different scenario to analyse potentials 
induced by massive vector particles.

Our object of interest is a neutral massive spin-1 mediating particle, which we might identify as a sort of massive photon. Such a particle is extensively discussed in the literature, dubbed as $Z^{0'}-$ particle. In the review articles 
of Ref. \cite{ref_Z_prime}, the authors present an exhaustive list of different $Z^{0'}-$ particles and phenomenological constraints on their masses and couplings. In our paper, we shall be studying interaction potentials between fermionic currents as induced by $Z^{0'}$ virtual particles; their effects are then included in the interpaticle potentials we are going to work out. Therefore, the velocity- and spin- dependence of our potentials appear as an effect of the interchange of a virtual $Z^{0'}-$ particle.

We exploit a variety of couplings to ordinary matter in order to extract possible experimental signatures 
that allow to distinguish between the two types of mediation in the regime of low-energy 
interactions. Just as in the usual electromagnetic case, where the 4-potential is subject to gauge-fixing 
conditions to reduce the number of degrees of freedom (d.o.f.), we shall also impose 
gauge-fixing conditions to the $\left\{A_{\mu}, B_{\nu \kappa} \right\}-$ system in order to ensure that only the spin-1 d.o.f. survives. From the physical side, we expect those potentials to exhibit a polynomial correction (in powers 
of $1/r$) to the well-known $e^{-m_{0}r}/r$ Yukawa potential. This implies that a laboratory aparatus with 
typical dimensions of $\sim mm$ could be used to examine the interaction mediated by 
massive bosons with $m_{0} \sim 10^{-3} eV$.

Developments in the measurement of macroscopic interactions between unpolarized and 
polarized objects \cite{Adelberger1}\cite{Adelberger2}\cite{Hunter}\cite{outras_aplicacoes} are able to 
constrain many of the couplings between electrons and nucleons (protons and neutrons), so 
that we can concentrate on more fundamental questions, such as the impact of the particular field 
representation of the intermediate boson in the fermionic interparticle potential. To 
this end, we discuss the case of monopole-dipole interactions in order to directly compare the Proca and $\left\{A_{\mu}, B_{\nu \kappa} \right\}$- mechanisms. We shall also present bounds on the vector/pseudo-tensor couplings that 
arise  from a possible application to the study of the hydrogen atom.
 

We would like to point out that our main contribution here is actually to associate
different field representations (which differ from each other by their respective 
off-shell d.o.f.) to the explicit spin-dependence in the particle potentials we derive. 
Rather than focusing on the constraints on the parameters, we aim at an understanding of 
the interplay between different field representations for a given spin and spin-spin 
dependence of the potentials that appear from the associate field-theoretic models. This shall be explicitly highlighted in the end of Section IV.B. We anticipate here however that
four particular types of spin- and velocity- dependences show up only in the topologically massive case we
discuss here. The Proca-type massive exchange do exclude these four terms, as it shall
become clear in Section IV.B.

Our paper is outlined according to what follows: in Section II, we introduce
 the concept of potential and briefly discuss the notation and conventions employed. 
Next, we calculate the potentials with different classes of couplings for the Proca 
and $\left\{A_{\mu}, B_{\nu \kappa} \right\}-$ system in Sections III and IV. 
In Section IV.A we present with due details the intermediate steps that yield the final expressions of
our set of propagators and we devote some words to compare our results with the propagators
worked out by other authors. We present our Conclusions and Perspectives in 
Section V. Two Appendices follow: in the Appendix A, we cast the list of relevant 
vertices in the low-energy limit. Next, in the Appendix B, we present the multiplicative 
algebra of a set of relevant spin operators that appear in the attainment of a set of 
propagators we have to compute in Section IV.


\section{Methodology}

Let us first establish the kinematics of our problem. We are dealing with two fermions, 1 and 2, which scatter elastically. If we work in the center of mass frame (CM), we can assign them momenta as indicated in Fig.\eqref{Fig1} below, where $\vec{q}$ is the momentum transfer and $\vec{p}$ is the average momemtum of fermion 1 before and after the scattering. 
\begin{figure}[h!]
\centering
\includegraphics[width=0.5\textwidth]{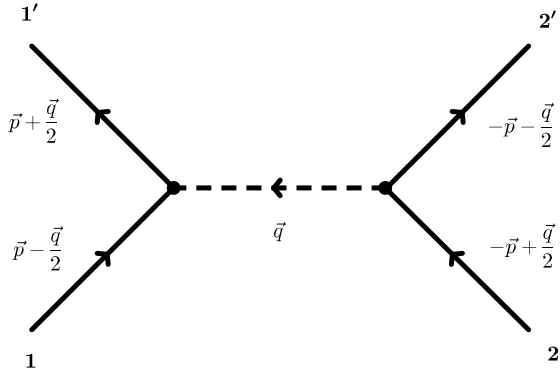}
\caption{\label{fig:overleaf}Basic vertex structure and momentum assignments.} \label{Fig1}
\end{figure}

Given energy conservation and our choice of reference frame, one can readily show that $\vec{p} \cdot \vec{q} = 0$ and that $q^{\mu}$ is space-like: $q^{2} = -\vec{q}^{\,2}$. The amplitude will be expressed in terms of $\vec{q}$ and $\vec{p}$ and we shall keep only terms linear in $|\vec{p}|/m_{1,2}$. It will also include the spin of the particles involved.

According to the first Born approximation, the two-fermion potential can be obtained from the Fourier transform of the tree-level momentum-space amplitude with respect to the momentum transfer $\vec{q}$
\begin{equation} 
V(r ,v) = - \int \frac{d^3\vec{q}}{(2\pi)^3} \, e^{i \vec{q} \cdot \,\vec{r}} \, \mathcal{A} ( \vec{q} , m \vec{v} ), 
\end{equation}
where $\vec{r}$, $r$ and $v = |\vec{p}|/m_{1,2}$ are the relative position vector, its modulus and average velocity of the fermions, respectively. The long-range behaviour is related to the non-analytical pieces of the amplitude in the non-relativistic limit \cite{Sucher}. We evaluate the fermionic currents up to first order in $|\vec{p}|/m_{1,2}$ and $|\vec{q}|/m_{1,2}$, as indicated in the Appendix A (an important exception is discussed in Section IV.B in connection with the mixed propagator $\langle A_{\mu} B_{\nu\kappa} \rangle$ since, in that case, contact terms arise).

We restrict ourselves to tree-level amplitudes since we are considering weakly interacting particles, thus carrying tiny coupling constants that suppress higher-order diagrams. The typical outcome are Yukawa-like potentials with extra $1/r$ contributions which also depend on the spin of the sources, as well as on their velocity. Contrary to the usual Coulomb case, spin- and velocity-dependent terms are the rule, not exception.


\section{The pure spin-1 case: the Proca field}

In order to establish the comparison between the two situations that involve a massive spin-1 particle, we start off by quickly reviewing the simplest realization of a neutral massive vector particle, the Proca field $A_{\mu}(x)$, described by the Lagrangean
\begin{equation} 
\mathcal{L}_{Proca} = - \frac{1}{4} \, F_{\mu \nu}^2 + \frac{1}{2} \, m_0^2 \, 
A_\mu^2 \label{L_proca} 
\end{equation}
with the field strength tensor given by $F_{\mu \nu} = \partial_\mu A_\nu - \partial_\nu A_\mu $. 

Since we are concerned with the interaction mediated by such a field, it is necessary to calculate its propagator, $\langle A_\mu A_\nu\rangle$. The Lagrangean above can be suitably rewritten as $\frac{1}{2} A^{\mu} \mathcal{O}_{\mu \nu} A^{\nu}$, in which the operator $\mathcal{O}_{\mu \nu}$, essentially the inverse of the propagator, is $\mathcal{O}_{\mu \nu} = \left( \Box + m_0^2 \right) \, 
\theta_{\mu \nu} + m_0^2 \omega_{\mu \nu}$, where we introduced the transverse and longitudinal projection operators defined as
\begin{eqnarray}
\theta_{\mu \nu} & \equiv & \eta_{\mu \nu} - \frac{\partial_\mu \partial_\nu}{\Box}, \\
\omega_{\mu \nu} & \equiv & \frac{\partial_\mu \partial_\nu}{\Box}, \label{projproc} 
\end{eqnarray}
which satisfy $\theta^{2} = \theta$, $\omega^{2} = \omega$, $\theta\omega = 0$ and $\theta + \omega = 1$. Due to these simple algebraic properties it is easy to invert $\mathcal{O}_{\mu \nu}$ and, transforming to momentum space, we finally have
\begin{equation} 
\langle A_\mu A_\nu\rangle = - \frac{i}{k^2 - m_0^2} \left( \eta_{\mu 
\nu} - \frac{ k_\mu k_\nu }{m_0^2} \right), \label{propproc} 
\end{equation}
from which we proceed to the calculation of the potentials.

Let us solve in more detail the case of two fermionic vector currents interacting via the Proca field. Using the parametrization of Fig.\eqref{Fig1} and applying the Feynman rules, we get 
\begin{eqnarray*} 
i \mathcal{A}_{V-V}^{Proca} & = & \bar{u}(p + q/2) \left\{ i g^V_1 
 \gamma^\mu  \right\} u(p - q/2) \langle A_\mu A_\nu\rangle \times \\ \nonumber
& \times & \bar{u}(- p - q/2) \left\{ i g^V_2 \gamma^\nu  \right\} 
u(- p + q/2)
\end{eqnarray*}
with $g^{V}_1$ and $g^{V}_2$ refering to the coupling constants. The equation above can be put in a simpler form as below
\begin{equation} 
\mathcal{A}_{V-V}^{Proca} =  i \, J_{1}^\mu \, \langle  A_\mu A_\nu \rangle \, 
J_{2}^\nu. 
\end{equation} 

If we use that $q^{0}=0$ and current conservation, we find that the amplitude is $\mathcal{A}_{V-V}^{Proca} =  -\frac{1}{\vec{q}^{\, 2} + m_0^2} J_1^\mu \, J_{2 \mu}$ and, according to eq.\eqref{vector_c_2}, we have $J_1^i \, J_{2i} \sim \mathcal{O}( v^2/c^2)$. Therefore, only the term $J^0_1 \, J_{20} \approx g^V_1 g^V_2 \delta_1 \delta_2$ contributes to the scattering amplitude, thus giving
\begin{equation} 
\mathcal{A}_{V-V}^{Proca} =  -g^{V}_1  g^{V}_2 \, \frac{\delta_1 
\delta_2}{\vec{q}^{\,2} + m_0^2}, \label{eq7}
\end{equation}
where $\delta_i$ ($i=1,2$ labels the particles) is such that $\delta_i=+1$ if the $i$-th particle experiences no spin flip in the interaction, and $\delta_i=0$ otherwise. In the eq.\eqref{eq7} above, the global term $\delta_1 \delta_2$ is present to indicate that the amplitude is non-trivial only if both particles do not flip their respective spins. If one of them changes its spin the potential vanishes. This means that this interaction only occurs with no spin flip. In what follows, we shall come across situations where only a single $\delta_i$ appears, thus justifying the effort to keep the $\delta_i$ explicit. 

Finally, we take the Fourier transform in order to obtain the potential between two static (vector) currents, 
\begin{equation} 
V^{Proca}_{V-V} =  \frac{g^{V}_1  g^{V}_2 \delta_1 \delta_2}{4 \pi}  \, \frac{e^{- m_0 r}}{r}, \label{eq8} 
\end{equation}
which displays the well-known exponentially suppressed repulsive Yukawa behaviour typical of a massive $s=1$ boson exchange. In our notation, the potential is indicated as $V_{v_{1} - v_{2}}$, where $v_{1,2}$ refer to the vertices related to the particles 1 and 2. In the case above, the subscripts $V$ stand for vector currents. As already announced, the typical decay length is $1/m_0$ and we expect that very light bosons will be measurable for (laboratory) macroscopic distances, e.g. for masses of $\sim 10^{-3} eV$, we have ranges of $d \sim mm$.

Following the same procedure, we can exploit other situations, namely: vector with pseudo-vector currents and two pseudo-vector currents. The results are cast in what follows:
\begin{eqnarray}
V^{Proca}_{V-PV} & = &  -\frac{g_1^V g_2^{PV}}{4 \pi } \Big\{  \vec{p} \,
\cdot \langle \vec{\sigma}\rangle_2  \, \frac{\delta_1}{r} \left[ \frac{1}{m_1} + 
\frac{1}{m_2} \right] +  \\ \nonumber
& + & \frac{\left( 1 + m_0 r \right)}{2 m_1 r^2} \, \left[\langle \vec{\sigma}\rangle_1 \times \langle \vec{\sigma}\rangle_2 \right] \cdot \hat{r} \,  \Big\} \, e^{-m_0 r} 
\end{eqnarray}
\begin{equation}
V^{Proca}_{PV-PV} =  -\frac{g_1^{PV} g_2^{PV}}{4 \pi } \, 
\langle \vec{\sigma}\rangle_1 \cdot \langle \vec{\sigma}\rangle_2  \, \frac{e^{-  m_0 
r}}{r},\end{equation}
and we notice that all kinds of spin-dependent interactions appear while the $r$ factors are limited to $r^{-2}$. It is also easy to see that $V^{Proca}_{PV-PV}$ and $V^{Proca}_{V-PV}$ are even and odd against a parity transformation, respectively. In the next section, we shall conclude that a richer class of potentials is generated if the massive spin-1 Abelian boson exhibits a gauge-invariant mass that comes from the mixing between a one- and a two-form potentials.


\section{The topologically massive spin-1 case}

The Proca vector field transforms under the $\left(\frac{1}{2},\frac{1}{2}\right)$-representation of the Lorentz group and its Lagrangean is the simplest extension leading to a massive intermediate vector boson, but it is not the only one. A massive spin-1 particle can also be described through a gauge-invariant formulation: a vector and a tensor fields connected by a mixing topological mass term \cite{Denis}. Both the vector $A_{\mu}$ and the tensor $B_{\mu\nu}$ are gauge fields described by the following Lagrangean:
\begin{equation}
\mathcal{L}_0 = - \frac{1}{4} \, F_{\mu \nu}^2 + \frac{1}{6} \, G_{\mu \nu \kappa 
}^2 + \frac{m_0}{\sqrt{2}} \,  \epsilon^{\mu \nu \alpha \beta} \, A_\mu \partial_\nu B_{\alpha \beta}, 
\label{L_B}
\end{equation}
where the field-strength for the anti-symmetric tensor is $G_{\mu \nu \kappa } = \partial_\mu B_{\nu \kappa } + \partial_\nu 
B_{ \kappa \mu} + \partial_\kappa B_{\mu \nu }$. The action is invariant under the independent local Abelian gauge transformations given by
\begin{eqnarray}
A'_\mu & = & A_\mu + \partial_\mu \alpha \\
B'_{\mu \nu} & = & B_{\mu \nu} + \partial_\mu \beta_\nu - \partial_\nu \beta_\mu, \label{Bgauge}
\end{eqnarray}
and it can be shown that together with the equations of motion, the pair $\{A_{\mu}, B_{\nu \kappa}\}$ carries three (on-shell) degrees of freedom, being, therefore, equivalent to a massive vector field. It is interesting to note that, contrary to the typical Proca case, the topological mass term does not break gauge invariance, so that no spontaneous symmetry breakdown is invoked.

Even though the Proca field and the mixed $\{A_{\mu}, B_{\nu \kappa}\}$-system describe both an on-shell spin-1 massive particle, these two cases are significantly different when considered off-shell. Our topologically massive spin-1 system displays 6 d.o.f. when considered off-shell (since gauge symmetry allows us to eliminate 4 compensating modes), whereas the Proca field carries 4 off-shell d.o.f. (the subsidiary condition, which is an on-shell statement, eliminates one d.o.f.). It is the on-shell spin-1 massive boson corresponding to the mixed $\{A_{\mu}, B_{\nu \kappa}\}$-system that we refer to as our $Z^{0'}-$type particle. Its exchange between external fermionic currents gives rise to the classes of interparticle potentials we wish to calculate and discuss in this paper.

On the other hand, since the potential evaluation is an off-shell procedure, we consider relevant to compare both situations bearing in mind that the potential profiles may indicate - if we are able to set up an experiment - whether a particular mechanism is preferable in the case of a specific physical system. Characteristic aspects of the potentials in these two situations might select one or other mechanism in some possible physical scenario, therefore being able to distinguish between different BSM models.

Our goal is to investigate the potentials between fermions induced by the exchange of the mixed vector and tensor fields and compare the spin-, velocity- and distance-dependence against the Proca case. To do that, we need, first of all, to derive the whole set of propagators.


\subsection{The propagators}

As in Section III, it is important to obtain suitable spin operators in order to 
obtain the propagators of the model. The spin operators that act on an 
anti-symmetric 2-form are

\begin{equation} 
\left( P^1_b \right)_{\mu \nu  , \, \rho \sigma } \equiv \frac{1}{2} 
\left( 
\theta_{\mu \rho} \, \theta_{\nu \sigma} - \theta_{\mu \sigma} \, \theta_{\nu \rho}
\right) \label{proj_B_1} 
\end{equation}
\begin{equation} 
\left( P^1_e \right)_{\mu \nu  , \, \rho \sigma } \equiv \frac{1}{2} 
\left( \theta_{\mu \rho} \, \omega_{\nu \sigma} + \theta_{\nu \sigma} \, \omega_{\mu \rho} -
\theta_{\mu \sigma} \, \omega_{\nu \rho} - \theta_{\nu \rho} \, \omega_{\mu \sigma}
\right) \label{proj_B_2} 
\end{equation}
which are anti-symmetric generalizations of 
the projectors $\theta_{\mu\nu}$ and $\omega_{\mu\nu}$ 
\cite{Helayel}\cite{proj_proca_1}\cite{proj_proca_2}. The comma indicates that we have anti-symmetry 
in changes $\mu \leftrightarrow \nu$ or $ \rho \leftrightarrow \sigma$. The algebra 
fulfilled by these operators is collected in the Appendix B. We quote them since they 
are very useful in the extraction of the propagators from Lagrangean \eqref{L_B}. Adding up the gauge-fixing terms to the Lagrangean \eqref{L_B},

\begin{equation} 
\mathcal{L}_{g.f.} = \frac{1}{2 \alpha} \, \left( \partial_\mu A^\mu \right)^2 + 
\frac{1}{2 \beta} \, \left( \partial_\mu B^{\mu \nu} \right)^2 \, ,\end{equation}
yields the full Lagrangean $\mathcal{L} = \mathcal{L}_0 + \mathcal{L}_{g.f.}$. In terms 
of the spin
operators, $\mathcal{L}$ can be cast in a more compact form as:

\begin{equation} 
\mathcal{L} = \frac{1}{2} \, \left( \begin{array}{cc} A^\mu & B^{\kappa \lambda }  
 \end{array}\right) \left( \begin{array}{cc} P_{\mu \nu} & Q_{\mu \rho \sigma }  \\
R_{\kappa \lambda \nu } & \mathbf{S}_{\kappa \lambda , \, \rho \sigma } 
\end{array}\right)
\left( \begin{array}{c} A^\nu \\ B^{\rho \sigma}  \end{array} \right),
\label{matrix_op} 
\end{equation}
where we identify
\begin{eqnarray} 
P_{\mu \nu} & \equiv & \Box \theta_{\mu \nu} - \frac{\Box}{\alpha} 
\omega_{\mu \nu} \\
Q_{\mu \rho \sigma } & \equiv & m_0 \, S_{\mu \rho \sigma}/\sqrt{2} \\
R_{\kappa \lambda \nu } & \equiv & - m_0 \, S_{\kappa \lambda \nu }/\sqrt{2} \\
\mathbf{S}_{\kappa \lambda , \, \rho \sigma } & \equiv & - \Box \left( 
P_b^1 \right)_{\kappa 
\lambda , \, \rho \sigma } - \frac{\Box}{2 \beta} \left( P_e^1 \right)_{\kappa \lambda 
, \, \rho \sigma }. 
\end{eqnarray}

With the help of Appendix B, we invert the matrix operator in $(\ref{matrix_op})$ 
and read off the $ \langle A_\mu A_\nu \rangle$, $\langle A_{\mu} B_{\kappa \lambda 
}\rangle $ and $\langle B_{\mu \nu} B_{\kappa \lambda }\rangle $ momentum-space 
propagators, which turn out to be given as below: 

\begin{equation} 
\langle A_\mu A_\nu\rangle = - \frac{i}{k^2 - m_0^2} \, \eta_{\mu \nu} + i\left( 
\frac{1}{k^2 - m_0^2}  +  \frac{\alpha}{k^2} \right) \, \frac{k_\mu 
k_\nu}{k^2} \label{prop_AA} 
\end{equation}
\begin{equation} 
\langle B_{\mu \nu} B_{\kappa \lambda }\rangle =  \frac{i}{k^2 -  m_0^2} \, \left( 
P_b^1 \right)_{\mu \nu , \, \kappa \lambda } + \frac{ 2i \beta}{k^2} \, \left( P_e^1 
\right)_{\mu \nu , \, \kappa \lambda } \label{prop_BB} 
\end{equation}
\begin{equation} 
\langle A_\mu B_{\nu \kappa }\rangle =  \frac{m_{0}/\sqrt{2}}{k^2 \left( k^2 -  m_{0}^{2} 
\right)} \, \epsilon_{\mu \nu \kappa \lambda } \, k^\lambda \label{prop_AB}. 
\end{equation}

From the propagators above, we clearly understand that the massive pole $k^2= m_0^2$, 
present in $(\ref{prop_AA})$-$(\ref{prop_AB})$, actually describes the spin-1 massive
excitation carried by the set $\{A_{\mu}, B_{\nu \kappa}\}$. In contrast to the off-shell regime of the so-called BF-model \cite{BF_model}, our non-diagonal $\langle A_\mu B_{\nu \kappa }\rangle$-propagator exhibits a massive pole 
and it cannot be considered separately from the $\langle A_\mu A_{\nu }\rangle-$ 
and $\langle A_{\mu} B_{ \kappa \lambda}\rangle$-propagators: only the full set of 
fields together correspond to the 3 d.o.f. of the on-shell massive spin-1 boson we 
consider in our study.

Different from the point of view adopted in Ref. \cite{Allen_et_al}, where the authors treat
the topological mass term as a vertex insertion (they keep the $ \langle A_\mu B_\nu \rangle -$ and
$ \langle B_{\mu \nu} B_{\kappa \lambda} \rangle -$ propagators separately and with a trivial pole $k^2 =
0$), we consider it as a genuine bilinear term and include it in the sector of 2-point functions. For that, we introduce the mixed spin operator $ S_{\mu \nu \kappa} $ in the algebra of operators and its final effect is to yield the mixed $ \langle A_\mu B_{\nu \kappa} \rangle -$propagator. The commom pole at $k^2 = m_{0}^2$ does not describe different particles, but a single
massive spin-1 excitation described by the combined $ \left\{ A_\mu , \, B_{\nu \kappa} \right\}-$ fields, as already stated in the previous paragraph. Ref. \cite{Allen_et_al} sums up the (massive) vertex insertions into the $ \langle A_\mu A_\nu \rangle -$ propagator which develops a pole at $k^2 = m^2$. They leave the $ \langle B_{\mu \nu} B_{\kappa \lambda} \rangle -$
propagator "aside" because the $B_{\mu \nu} -$ field does not interact with the fermions; the latter
are minimally coupled only to $A_\mu$.

On the other hand, in Ref. \cite{Leblanc_et_al}, the topological mass term that mixes
$A_\mu$ and $B_{\nu \kappa}$ is generated by radiative corrections induced by the 4-fermion
interactions. So, for the sake of their calculations, the authors work with a massless vector
propagator whose mass is dynamically generated. This is not what we do here. In a more recent paper \cite{Diamantini_et_al}, again an induced topological mass term mixes $A_\mu$ and $ B_{\nu \kappa} $ but, in this case, it is a topological current that radiatively generates the
mass. 

We point out the seminal paper by Cremmer and Scherk \cite{CS}, where they show that, for the spectrum analysis, it is possible to take the field-strength $G_{\mu \nu \kappa}$ and its dual $\widetilde{G}_\mu$, as fundamental fields, thus enabling them to go into a new field basis where a Proca-like field emerges upon a field redefinition. We cannot follow this road here, for our $B_{\mu \nu}$  is coupled to a tensor and to a pseudo-tensor currents in the process of evaluating some of our potentials. This prevents us from adopting $\widetilde{G}_\mu$ as a fundamental field, as it is done in \cite{CS}; this would be conflicting with the locality of the action. But, for the sake of analysing the spectrum, Cremmer and Scherk's procedure works perfectly well. 

Finally, we also point out the paper by Kamefuchi, O' Raifeartaigh and Salam \cite{Salam_Kamefuchi} that discusses the conditions on field reshufflings which do not change the physical results, namely, the $S-$matrix elements. A crucial requirement is that the change of basis in field space do not yield non-local interactions. 

To conclude the present sub-section on the propagators, we reinforce that once the $A_{\mu}$- and $B_{\mu\nu}$-fields interact with external currents, the diagonalization of the (free) bilinear piece of the Lagrangean is not a good procedure, the reason being that the topological mass term has a derivative operator, which would imply into non-local interactions between the new (diagonalized) fields and the external currents, so that the physical equivalence stated in the Kamefuchi-O'Raifeartaigh-Salam's paper can no longer be undertaken.


\subsection{The potentials}

We have already discussed the procedure to obtain the spin- and velocity-dependent potentials in previous 
sections. Thus, we shall focus on the particular case in which we have the 
propagator $ \langle B_{\mu \nu} B_{\kappa \lambda }\rangle $ and two 
tensor currents. In the following, we adopt the same parametrization of Fig.\eqref{Fig1}. After applying the 
Feynman rules, we can rewrite the scattering amplitude for this process as
\begin{equation} 
\mathcal{A}^{\langle BB\rangle}_{T-T} = i J^{\mu \nu}_1 \langle B_{\mu \nu} B_{\kappa 
\lambda }\rangle J_2^{\kappa \lambda } \label{amp_BB_TT} 
\end{equation}
with the tensor currents given by eq.\eqref{tensor_c_0}. Substituting the propagator $(\ref{prop_BB})$ in eq.$(\ref{amp_BB_TT})$ and eliminating its longitudinal sector (due to current conservation), we have $\mathcal{A}^{\langle BB\rangle}_{T-T} =  -\frac{1}{q^2 - m_0^2} J_1^{\mu \rho } J_{2 \mu \rho }$. The product of currents leads to $J_1^{\mu \rho } J_{2 \mu \rho } = 2 J_1^{0i } J_{2 \,0i } + J_1^{ij} J_{2 \, ij }$. However, according to eq.$(\ref{tensor_c_1})$, we conclude 
that $J_1^{0i } J_{2 \,0i } \sim \mathcal{O}(v^2/c^2)$ does not contribute to the 
non-relativistic amplitude. The term $J_1^{ij} J_{2 \, ij }$ can be simplified by using eq.$(\ref{tensor_c_2})$ (with the appropriate changes to the second current), so that we get $\mathcal{A}^{\langle BB\rangle}_{T-T} = \frac{1}{2} \, \frac{g^T_1 
g^T_2}{\vec{q}^{\, 2} + m_0^2} \, \langle \vec{\sigma}\rangle_1 \cdot \langle \vec{\sigma}\rangle_2$. Performing the well-known Fourier integral, we obtain the non-relativistic spin-spin potential, namely
\begin{equation} 
V^{\langle BB\rangle}_{T-T}=  - \frac{g_1^T g_2^T}{8 \pi} \, \langle \vec{\sigma}\rangle_1 \cdot 
\langle \vec{\sigma}\rangle_2 \frac{e^{- m_0 r}}{r},\label{Pot_T_T} 
\end{equation}
and, similarly, we find the interaction potentials between tensor and pseudo-tensor currents  currents to be
\begin{eqnarray}
V^{\langle BB\rangle}_{T-PT} & = &  \frac{g_1^T g_2^{PT}}{8 \pi r} \, \Big\{ \left(\frac{1}{m_{1}} + \frac{1}{m_{2}}\right) \, \vec{p} \cdot \left( \langle \vec{\sigma}\rangle_1 
\times \langle \vec{\sigma}\rangle_2 \right) + \\ \nonumber
& + & \frac{\left( 1 + m_0 r \right)}{2r} \, \left( \frac{\delta_2}{m_2}  \langle \vec{\sigma}\rangle_1 - 
\frac{\delta_1}{m_1} \langle \vec{\sigma}\rangle_2  \right) 
\cdot \hat{r} \,  \Big\} \, e^{- 
m_0 r}
\end{eqnarray}
as well as two pseudo-tensors
\begin{equation} 
V^{\langle BB\rangle}_{PT-PT} = \frac{g_1^{PT} g_2^{PT}}{8 \pi} \, 
\langle \vec{\sigma}\rangle_1 
\cdot \langle \vec{\sigma}\rangle_2 \frac{e^{-m_0 r}}{r}.\label{Pot_PT_PT}
\end{equation}

It is worthy comparing the potentials $(\ref{Pot_T_T})$ and $(\ref{Pot_PT_PT})$. We observe that they differ by a relative minus sign. This means that they exhibit opposite behaviors for a given spin configuration: one is attractive and the other repulsive. The physical reason is that the $PT-PT$ and $T-T$ potentials stem from different sectors of the currents: the $PT-PT$ amplitude is composed by the $(0i)-(0j)$ terms of the currents; the $T-T$ amplitude, on the other hand, arises from the $(ij)-(kl)$ components, as it can be seen from eq.\eqref{amp_BB_TT}. 

In the light of that, we check the structure of the $\langle B_{\mu \nu} B_{\kappa \lambda }\rangle$-propagator and it becomes clear that, in the case of the $\langle B_{0i} B_{0j}\rangle$-mediator, an off-shell scalar mode is exchanged. In contrast, in the $\langle B_{ij} B_{kl}\rangle$-sector the only exchange is of a pure $s=1$ (off-shell) quantum. It is well-known, however, that the exchange of a scalar and a $s=1$ boson between sources of equal charges yields attractive and repulsive interactions, respectively, therefore justifying the aforementioned sign difference between eqs.$(\ref{Pot_T_T})$ and $(\ref{Pot_PT_PT})$.

For the mixed propagator $\langle A_\mu B_{\kappa \lambda }\rangle$, eq.\eqref{prop_AB}, we have 
four possibilities envolving the following currents: 
vector with tensor, vector with pseudo-tensor, pseudo-vector with 
tensor and pseudo-vector with pseudo-tensor. The results are given below:

\begin{eqnarray}
V^{\langle AB\rangle}_{V-T} & = & \frac{ g_1^{V} g_2^{T} \delta_1}{4\pi\sqrt{2}  m_0 r^2} \, 
\left[ 1 - \left( 1 +  m_0 r \right)  \, 
e^{-m_0 r} \right]\langle \vec{\sigma}\rangle_2 \cdot \hat{r} \\
& & \nonumber \\
V^{\langle AB\rangle}_{PV-T} & =  & \frac{g_1^{PV}g_2^{T}}{4\pi\sqrt{2}m_0 \mu r^2} \big[1 - \left( 1  
 + m_0 r \right)e^{-m_0 r}\big]\left( \langle \vec{\sigma}\rangle_1 
\cdot \vec{p} \right)\left( \langle \vec{\sigma}\rangle_2 \cdot \hat{r} \right) \\
& & \nonumber \\
V^{\langle AB\rangle}_{PV-PT} & = & \frac{ g_1^{PV} g_2^{PT}}{\sqrt{2} m_0} \biggl\{  \frac{\delta_{2}}{2m_{1}m_{2}}\left[\delta^3(\vec{r}) + \frac{m_0^2}{4 \pi r}e^{-m_0 r}\right] \langle \vec{\sigma} \rangle_{1}\cdot \vec{p}  + \nonumber \\  
& + &  \frac{1}{4\pi r^{2}} \left[ 1 - \left( 1 + m_0 r \right)e^{-m_0 r} \right] \left(\langle \vec{\sigma}\rangle_2 \times \langle \vec{\sigma}\rangle_1 \right) \cdot \hat{r}   \biggr\}.
\end{eqnarray}

The richest potential is the one between vector and pseudo-tensor sources, given by

\begin{eqnarray}
V_{V-PT}^{^{\langle AB\rangle}} & = & \frac{g^{V}_{1}g^{PT}_{2}}{\sqrt{2}m_{0}} \biggl\{ \frac{\delta_{1} 
\delta_{2}}{2m_{2}} \left[\delta^3(\vec{r}) + \frac{m_{0}^{2}}{4\pi r} e^{-m_{0}r} \right] + \nonumber \\ 
& + & \frac{\delta_{1}}{4 \pi \mu r^{3}}  \left[1 - \left(1 + m_{0}r\right)e^{-m_0 r}\right]\vec{L} \cdot \langle \vec{\sigma}\rangle_2 + \nonumber \\
& + & \frac{1}{2m_{1}}\left[\delta^3(\vec{r}) + \frac{m_{0}^{2}}{4\pi r}e^{-m_{0} r} - \frac{1}{4\pi r^{3}}\left[ 1 + \left( 1 + m_{0}r \right)e^{-m_{0}r} \right] \right]\langle \vec{\sigma}\rangle_1 \cdot \langle \vec{\sigma}\rangle_2 + \nonumber \\
& + & \frac{1}{8\pi m_{1} r^{3}}\left[ 3 + \left( 3 + 3m_{0}r + m_{0}^{2}r^{2} \right)e^{-m_{0}r} \right] \left( \langle \vec{\sigma}\rangle_1 \cdot \hat{r} \right) \left( \langle \vec{\sigma}\rangle_2 \cdot 
\hat{r} \right) \biggr\} \label{LS}
\end{eqnarray}
where we have introduced the reduced mass of the fermion system $\mu^{-1} = m_{1}^{-1} + m_{2}^{-1}$ and $\vec{L} = \vec{r} \times \vec{p}$ stands for the orbital angular momentum. 

Naturally, the contact terms do not contribute to a macroscopic interaction. Nevertheless, they are significant in quantum-mechanical applications in the case of {\it s-waves} which can overlap the origin. This is a peculiarity of $\langle A_\mu B_{\kappa \lambda }\rangle$-sector due to the extra $q^{2}$-factor in the denominator, which forces us to keep terms of order $|\vec{q}|^{2}$ in the current products. 

For the propagator $\langle A_\mu A_\nu\rangle$, eq,$(\ref{prop_AA})$, we find the 
same results as the ones in the Proca situation, due to current conservation. This means that, even though the vector field appears now mixed with the $B_{\mu \nu}$-field with a gauge-preserving mass term, for the sake of the interaction potentials, the results are the same as in the Proca case as far as the $A_\mu$-field exchange is concerned. 

We mention in passing that the $V^{\langle BB\rangle}_{T-T}$, $V^{\langle BB\rangle}_{PT-PT}$, $V^{\langle AB\rangle}_{PV-T}$ and $V^{\langle AB\rangle}_{V-PT}$ potentials are even under parity, while $V^{\langle BB\rangle}_{T-PT}$, $V^{\langle AB\rangle}_{V-T}$ and $V^{\langle AB\rangle}_{PV-PT}$ are odd. This difference is due to the presence of a single factor of the momentum transfer in the mixed propagator, eq.\eqref{prop_AB}.

We point out that experiments with rare earth iron garnet test masses \cite{Leslie} could be a possible scenario to distinguish  the two different mass mechanisms. In the Proca mechanism, we obtained the following spin- and velocity-dependences: 
$ \vec{p} \cdot \vec{\sigma} \, $, $ \,( \vec{\sigma}_1  \times \vec{\sigma}_2   ) \cdot \hat{r} $
and $\vec{\sigma}_1 \cdot \vec{\sigma}_2 $. These also appear in the gauge-preserving mass, but there we have additional profiles, given by $( \vec{\sigma}_1  \times \vec{\sigma}_2   ) \cdot \vec{p} $,
$\, \vec{\sigma} \cdot \hat{r} $, $ \, ( \vec{\sigma}_1  \cdot \vec{p}   )  ( \vec{\sigma}_2  \cdot \hat{r}   ) $ and
$\,  ( \hat{r}  \times \vec{p}   ) \cdot \vec{\sigma} $. 

The experiment provides six configurations $(C1,...,C6)$ by changing the relative orientation of the detector and the test mass (with respective spin polarizations and relative velocities). One of these configurations is interesting to our work, namely, the $C5$ is sensitive only to $\,  ( \hat{r}  \times \vec{p}   ) \cdot \vec{\sigma} $ dependence, which is only present in the gauge-preserving mass mechanism. For the other profiles we cannot distinguish the contributions of different mechanisms in this experiment. For example, the $C2$ configuration is sensitive to both $ \, ( \vec{\sigma}_1  \cdot \vec{p} )  ( \vec{\sigma}_2  \cdot \hat{r}) $ and $ \vec{\sigma}_1  \cdot \vec{\sigma}_2 $ dependences.


\section{Conclusions and perspectives}

The model we are investigating describes an extra Abelian gauge boson, a sort of $Z^{0'}$, which appears as a neutral massive excitation of a mixed $\{A_\mu, B_{\nu \kappa}\}-$ system of fields. It may be originated from some sector of BSM physics, where the coupling between an Abelian field and the 2-form gauge potential in the SUGRA multiplet may yield the topologically massive spin-1 particle we are considering. To have detectable macroscopic effect, this intermediate particle should have a very small mass, of the order of meV. This would be possible in the class of phenomenological models with the so-called large extra dimensions.

It is clear that the considerable number of off-shell degrees of freedom of the $\{A_\mu, B_{\nu \kappa}\}-$ model accounts for the variety of potentials presented above. In order to distinguish between the two models, a possible experimental set-up could consist of a neutral and a polarized source (1 and 2, respectively). Suppose, furthermore, that the sources display all kinds of interactions (V, PV, T, etc). 

\begin{figure}[h!]
\centering
\includegraphics[width=0.55\textwidth]{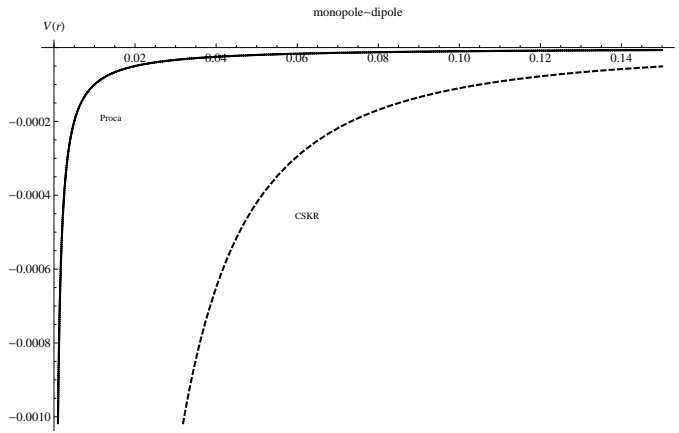}
\caption{\label{fig:overleaf}Monopole-dipole potentials with $m_{1}=m_e = 10^{5} eV$, $m_{0}=10^{-3} eV$ and source 1 velocity of order $v \simeq 10^{-6}$. The scale is irrelevant and coupling constants were not included for simplicity.} \label{Fig2}
\end{figure}

In this case, we must collect the terms proportional to $\langle \vec{\sigma}\rangle_2 \equiv \langle \vec{\sigma}\rangle$ in the two scenarios
\begin{eqnarray}
V^{Proca}_{mon-dip} &=& -\frac{g^{2}}{\mu} \frac{e^{-m_{0}r}}{r} \vec{p}\cdot \langle \vec{\sigma}\rangle 
\end{eqnarray}
and
\begin{eqnarray}
V^{\left\{A,B\right\}}_{mon-dip} & = & -\frac{g^{2} }{\mu} \frac{e^{-m_{0}r}}{r} \vec{p}\cdot \langle \vec{\sigma}\rangle + \nonumber \\
& - & \frac{g^{2}}{m_{1}} \frac{(1 + m_{0}r)e^{-m_{0}r}}{r^{2}} \hat{r} \cdot \langle \vec{\sigma}\rangle + \nonumber \\
& + & \frac{g^{2}}{m_{0}} \frac{\left[ 1 - \left(1 +  m_{0}r \right)e^{-m_{0}r} \right]}{r^{2}}
 \hat{r} \cdot \langle \vec{\sigma}\rangle + \nonumber \\
& - & \frac{g^{2}m_{0}}{m_{1}m_{2}} \frac{e^{-m_{0}r}}{r} \vec{p} \cdot \langle \vec{\sigma}\rangle \nonumber \\
& + & \frac{g^{2} }{\mu m_{0}} \frac{\left[1 - \left(1 + m_{0}r\right)e^{-m_0 r}\right]}{r^{3}} \left(\vec{r} \times \vec{p}\right) \cdot \langle \vec{\sigma}\rangle, \label{md}
\end{eqnarray}
where, for simplicity, we have omitted the labels in the coupling constants. In the macroscopic limit these would be effectively substituted by $g \rightarrow gN_{i}$, being $N_{i}$ the number of interacting particles of type $i$ in each source. If we consider the case in which the source 1 carries momentum so that $\vec{p} \,  // \, \langle \vec{\sigma} \rangle$, the last term above vanishes. Similarly, it is easy to see that the third term is essencially constant, while the fourth one is negligeable, since $m_{0}|\vec{p}|/m_{1}m_{2} \ll 1$ by definition. In Fig.\eqref{Fig2}, we plot the two resulting potentials.

It would then be possible, in principle, to determine which field representation, Proca or $\{A_{\mu}, B_{\nu\kappa}\}$, better describes the interaction at hand. It is worth mentioning that this difference is regulated by the $1/m_1$ factor in the second term of eq.\eqref{md}, so that only the lightest fermions (i.e., electrons and not the protons or neutrons, provided that, in a macroscopic source, we can safely neglect the internal structure of the nucleons) would be able to contribute significantly. 


The calculation we have performed is based on the quantum field-theoretical scattering amplitude in the non-relativistic limit, and the potential obtained - which can be interpreted as an operator - is also suitable to be introduced in the Schr\"odinger equation as a time-independent perturbation to the full Hamiltonian. This is a reasonable approach if these corrections are relatively small, which is to be expected, given that the standard quantum mechanical/QED results are in good agreement with experiments.

If we take the second line of eq.\eqref{LS}, for example, we notice a coupling of the angular momentum of the first fermion with the spin of the second. Such a spin-orbit coupling is also found in the hydrogen atom, contributing to its fine structure (with typical order of magnitude of $10^{-6¨} eV$). Supposing that the proton and electron are charged under the gauge symmetries leading to the $\left\{A_\mu B_{\nu \kappa}\right\}-$ fields, we can calculate a correction to the energy levels of their bound state due to $\langle A_\mu B_{\kappa \lambda }\rangle$ exchange as a means of estimation for the $V-PT$ coupling constants as a function of $m_{0}$.
Expanding the exponential in $1-(1+m_{0}r)e^{-m_{0}r}$ and keeping only the leading term, the spin-orbit term simplifies to
\begin{eqnarray}
V_{V-PT}^{LS} & = & \frac{\sqrt{2}g^{V}_{1}g^{PT}_{2} m_{0}}{8\pi\mu}\frac{1}{r} {\bf L}\cdot{\bf S}
\end{eqnarray}
with ${\bf S}=\langle \vec{\sigma}\rangle_2 /2$. Applying first-order perturbation theory to this potential gives a correction to the energy of $\Delta E^{LS} = \frac{g^{V}_{1}g^{PT}_{2} m_{0}}{8\pi\sqrt{2}\mu (n^{2}a_{0})} X_{l}$, where $X_{l} = l$ for $j=l+1/2$ and $X_{l} = -(l+1)$ for $j=l-1/2$. As we are interested in an estimate, we suppose $|X_{l}|/n^{2} \sim 1$. Given that the reduced mass and the Bohr radius are $\mu \simeq m_{e} = 5.11 \times 10^{5}$ eV and $a_{0} = 2.69 \times 10^{-4}$ $eV^{-1}$, respectively, we can constrain $\Delta E^{LS}$ to be smaller than the current spectroscopic uncertainties of one part in $10^{14}$ \cite{Parthey}. We then obtain $|g^{V}g^{PT}| < 10^{-8}$, for a mass of order $m_{0} \sim 10^{-2} eV$, which poses a less stringent, but consistent (in regard to the orders of magnitude of other couplings \cite{Dobrescu}), upper bound on the couplings. We see that this correction is much smaller than the typical spin-orbit contribution. A more comprehensive study applying atomic spectroscopy of both electronic and muonic hydrogen atoms will be reported in a forthcoming paper.

At last, but not less interesting, we indicate that it is possible to assign certain $CP$-transformation properties to the fields $A_{\mu}$ and $B_{\mu\nu}$ so that the topological mass term in eq.\eqref{L_B} violates $CP$. This would induce an electric dipole moment (EDM) if we couple our model to fermionic fields. Following the procedure employed by Mantry {\it et al} \cite{Mantry} in the context of axions, one could also use information from the EDM to find further bounds on the coupling constants and the mass of the intermediate spin-1 boson.


\begin{acknowledgments}
We would like to thank the National Council for Scientific and Technological Development of Brazil (CNPq) and the Co-ordination for Qualification of Higher Education Personnel (CAPES) for the invaluable financial support.
\end{acknowledgments}


\appendix

\section{Currents in the non-relativistic approximation}

\indent

In the following we present a brief summary of the conventions and main decompositions 
employed in the calculations carried out in the previous Sections.


\subsection{Basic conventions}

\indent

The basic spinors used to compose the scattering amplitude are the positive energy solutions to the Dirac equation in momentum space \cite{Ryder}, namely
\begin{equation} 
u(p) = \left( \begin{array}{c} \xi \\ \frac{\vec{\sigma} \cdot 
\vec{p}}{2m} \, \xi\end{array} \right) 
\end{equation}
where $ \xi = \left( \begin{array}{c} 1 \\ 0\end{array} \right) $ or 
$ \xi = \left(\begin{array}{c} 0 \\ 1 \end{array} \right) $ for spin-up and -down, respectively. Above we have assumed the non-relativistic limit $E + m \approx 2m$. The orthonormality relation $\xi'^\dagger _r \xi_s = \delta_{rs}$ is supposed to hold and we will usually suppress spinor indices. 

The gamma matrices are chosen as
\begin{equation} 
\gamma^0 = \left( \begin{array}{cc} 1 & 0 \\ 0 & -1 \end{array} \right) 
\, \; \, \text{and} \; 
\, \; \gamma^i = \left( \begin{array}{cc} 0 & \sigma_i \\ - \sigma_i & 0 \end{array} 
\right),
\end{equation}
and the metric and Levi-Civita symbol are defined so that $\eta^{\mu \nu} = \textrm{diag}(+,-,-,-)$ and $\epsilon^{0123} = +1$, respectively. We adopt natural units $\hbar = c = 1$ throughout.


\subsection{Current decompositions}

\indent

In order to calculate the spin-dependent potentials, it is useful to have the 
non-relativistic limit of the source currents, where we assume 

\begin{enumerate}
\item[{\bf 1)}] $ |\vec{p}|^{2}/m^2  \sim  \mathcal{O}\left(v^2 \right) 
\, \rightarrow \, 0$
\item[{\bf 2)}] Small momentum transfer: $ |\vec{q}|^{2}/m^2 
\, \rightarrow \, 0$
\item[{\bf 3)}] The cross product tends to zero if $|\vec{p}|/m $ and $|\vec{q}|/m$ are small. Energy-momentum conservation implies $\vec{p} \cdot \vec{q} = 0$ 
\end{enumerate}

Here, we show the results of the main 
fermionic currents. We adopt the parametrization for the first current (i.e., first vertex), 
following Fig.\eqref{Fig1}. We denote the generators of the boosts and rotations by
\begin{equation} 
\Sigma^{\mu \nu} \equiv - \frac{i}{4} \, \left[ \gamma^\mu , \gamma^\nu 
\right],
\end{equation}
and $\langle \sigma_i\rangle \,\equiv \xi'^\dagger \, \sigma_i \, \xi$. In the Dirac representation, $\gamma_5$ is given by
\begin{equation} 
\gamma_5 = \left( \begin{array}{cc}  0 & 1 \\ 1 & 0 \end{array} \right). 
\end{equation}

Making use of the Dirac spinor conjugate, $\bar{u} \equiv u^\dagger \gamma^0$, we have the following set of identities, omitting the coupling constants: 

\begin{enumerate}

\item[1)] Scalar current $(S)$:
\begin{equation} \bar{u}(p + q/2) \, u(p - q/2) \approx \delta \, . \label{scalar_c} 
\end{equation}


\item[2)] Pseudo-scalar current $(PS)$:

\begin{equation}  
\bar{u}(p + q/2) \, i \gamma_5 \, u(p - q/2) =  
- \frac{i}{2m} \, \vec{q} \cdot \langle \vec{\sigma}\rangle\,
 \label{pseudo_scalar_c}  
 \end{equation}


\item[3)] Vector current $(V)$:
\begin{equation} 
\bar{u}(p + q/2) \, \gamma^\mu \, u(p - q/2), 
\end{equation}

\begin{enumerate}


\item[{\bf 3i)}] For $\mu = 0$,

\begin{equation} 
\bar{u}(p + q/2) \, \gamma^0 \, u(p - q/2) \approx  \delta 
\label{vector_c_1} 
\end{equation}
\item[{\bf 3ii)}] For $\mu = i$,
\begin{equation} 
\bar{u}(p + q/2) \, \gamma^i \, u(p - q/2) = 
 \frac{\vec{p}_i}{m} \, \delta - \frac{i}{2m} \, \epsilon_{ijk} \, \vec{q}_j \, 
 \langle \sigma_k \rangle  \label{vector_c_2} 
 \end{equation}
 
\end{enumerate}


\item[4)] Pseudo-vector current $(PV)$:
 \begin{equation} 
 \bar{u}(p + q/2) \gamma^\mu \gamma_5  u(p - q/2)  
 \end{equation}
 
\begin{enumerate}

\item[{\bf 4i)}] For $\mu = 0$,

\begin{equation} 
\bar{u}(p + q/2) \, \gamma^0 \, \gamma_5 \, u(p - q/2) =
  \frac{1}{m} \, \langle \vec{\sigma}\rangle \cdot \vec{p}   
 \label{pseudo_vector_c_1} 
 \end{equation}
 
\item[{\bf 4ii)}] For $\mu = i$,
\begin{equation} 
\bar{u}(p + q/2) \, \gamma^i \, \gamma_5 \, u(p - q/2) \, \approx \,
  \langle \sigma_i\rangle   \label{pseudo_vector_c_2} 
\end{equation}

\end{enumerate}


\item[5)] Tensor current $(T)$:
\begin{equation} 
\bar{u}(p + q/2) \, \Sigma^{\mu \nu} \, u(p - q/2)  
\label{tensor_c_0} 
\end{equation}

\begin{enumerate}

\item[{\bf 5i)}] For $\mu = 0 \, $ and $\, \nu = i \,$, 

\begin{equation} 
\bar{u}(p + q/2) \, \Sigma^{0 i} \, u(p - q/2) = 
  \frac{1}{2m} \, \epsilon_{ijk} \, \vec{p}_j \, \langle \sigma_k\rangle  + 
\frac{i}{4m} \, \delta \, \vec{q}_i  \label{tensor_c_1}   
\end{equation}

\item[{\bf 5ii)}] For $\mu = i \,$ and $ \, \nu = j \,$,
\begin{equation} 
\bar{u}(p + q/2) \, \Sigma^{ij} \, u(p - q/2) \approx  
   - \frac{1}{2} \, \epsilon_{ijk} \langle \sigma_k\rangle    \label{tensor_c_2} \end{equation}
   
\end{enumerate}


\item[6)] Pseudo-tensor current $(PT)$:

\begin{equation} 
\bar{u}(p + q/2) \, i \, \Sigma^{\mu \nu} \, \gamma_5 \, u(p - q/2) 
\end{equation}

\begin{enumerate}

\item[{\bf 6i)}] For $\mu = 0 \, $ and $ \, \nu = i$,

\begin{equation} 
\bar{u}(p + q/2) \, i \, \Sigma^{0i} \, \gamma_5 \, u(p - q/2) \approx  
 \frac{1}{2} \langle \sigma_i\rangle  \label{pseudo_tensor_c_1}  
 \end{equation}
 
\item[{\bf 6ii)}] For $\mu = i \, $ and $ \, \nu = j $

\begin{eqnarray}
\bar{u}(p + q/2) \, i \, \Sigma^{ij} \, \gamma_5 \, u(p - q/2) & = &  \frac{1}{2m} \, \left( \vec{p}_i \langle \sigma_j\rangle - \vec{p}_j \langle \sigma_i\rangle \right)  + \\ \nonumber
& + & \frac{i}{4m} \, \delta \, \epsilon_{ijk}  \, \vec{q}_k  \label{pseudo_tensor_c_2} 
\end{eqnarray}  

\end{enumerate}
\end{enumerate}

In the manipulations above, we have kept the $rs$ indices implicit in the $\delta_{rs}$, as adopted in the main text, pointing out only the particle label. Due to momentum conservation and our choice of reference frame (CM), the second current (or second vertex) can be obtained by performing the changes $q \rightarrow - q$ and $p \rightarrow - p$ in the first one.


\section{Spin operators}

The spin operators satisfy the following algebra:

\begin{equation} 
\left( P^1_b + P^1_e \right)_{\mu \nu  , \, \rho \sigma } = \frac{1}{2} 
\left(\eta_{\mu \rho} \eta_{\nu \sigma} - \eta_{\mu \sigma} \eta_{\nu \rho} 
\right) \equiv 1_{\mu \nu  , \, \rho \sigma }^{a.s.} \end{equation}

\begin{equation} 
\left( P^1_b \right)_{\mu \nu  , \, \alpha \beta} 
\left( P^1_b \right)^{\alpha \beta}_{  \, \; \; \, , \, \rho \sigma} = 
\left( P^1_b \right)_{\mu \nu  , \, \rho \sigma} \end{equation} 

\begin{equation} 
\left( P^1_e \right)_{\mu \nu  , \, \alpha \beta} 
\left( P^1_e \right)^{\alpha \beta}_{  \, \; \; \, , \, \rho \sigma} = 
\left( P^1_e \right)_{\mu \nu  , \, \rho \sigma} \end{equation} 

\begin{equation} 
\left( P^1_b \right)_{\mu \nu  , \, \alpha \beta} 
\left( P^1_e \right)^{\alpha \beta}_{  \, \; \; \, , \, \rho \sigma} = 0 
\end{equation}

\begin{equation} 
\left( P^1_e \right)_{\mu \nu  , \, \alpha \beta} 
\left( P^1_b \right)^{\alpha \beta}_{  \, \; \; \, , \, \rho \sigma} = 0.
\end{equation}

We notice that the mixing term between $A_\mu$ and 
$B_{\mu \nu}$ introduces a new operator, $S_{\mu \nu \kappa} \equiv \epsilon_{\mu \nu \kappa \lambda} 
\, \partial^\lambda$, which is {\it not} a projector, since
\begin{equation} 
\epsilon^{\mu \nu \alpha \beta} \, A_\mu \partial_\nu B_{\alpha 
\beta} = \frac{1}{2} \left[A^\mu \, S_{\mu \kappa \lambda } B^{\kappa \lambda } 
-  B^{\kappa \lambda } \, S_{\kappa \lambda \mu } \, A^\mu \right], 
\end{equation}
so that we need to study the algebra of $S_{\mu \nu \kappa}$ with the projectors 
$(\ref{proj_B_1})$ and $(\ref{proj_B_2})$, giving us

\begin{equation} S_{\mu \nu \alpha} S^{\alpha \kappa \lambda} = - 2\Box 
\left( P^1_b \right)^{ \, \; \, \; \, \; \kappa \lambda}_{ \mu \nu ,} \end{equation}

\begin{equation} \left( P^1_b \right)_{\mu \nu , \, \alpha \beta } S^{\alpha \beta 
\kappa} = S_{\mu \nu}^{\; \, \; \, \; \kappa} \end{equation}

\begin{equation} S^{\kappa \alpha \beta} 
\left( P^1_b \right)_{\alpha \beta ,}^{  \, \; \; \, \; \, \; \mu \nu} = S^{\kappa \mu 
\nu} \end{equation}

\begin{equation} \left( P^1_e \right)_{\mu \nu , \, \alpha \beta } S^{\alpha \beta 
\kappa} = 0 \end{equation}

\begin{equation} 
S^{\kappa}_{\; \, \; \, \alpha \beta} 
\left( P^1_e \right)^{ \alpha \beta , \, \mu \nu } = 0 \end{equation}

\begin{equation} 
S_{\mu \alpha \beta} S^{\alpha \beta}_{\, \; \, \; \nu} = - 2 \Box 
\theta_{\mu \nu} \label{S_S_theta}.
\end{equation}

The possibility to obtain a closed algebra is not only desirable, but very important, in order to complete the inversion of the matrix in eq.\eqref{matrix_op}.


\end{document}